\documentclass{emulateapj}

\usepackage{natbib}
\usepackage{algorithm}

\usepackage[nooneline]{caption2}
\usepackage{algorithmic}
\usepackage{graphicx}
\usepackage{amsfonts}
\usepackage{amsmath}
\usepackage{amssymb,enumerate}

\usepackage[hyperfootnotes=false]{hyperref}

\shorttitle{Bayesian Time Symmetry Analysis}
\shortauthors{Blocker, Protopapas, \& Alcock}

\begin{document}

\title{A Bayesian approach to the analysis of time symmetry in light curves: \\
Reconsidering Scorpius X-1 occultations}

\author{Alexander W. Blocker\altaffilmark{1}, Pavlos Protopapas
\altaffilmark{2,3} and Charles R. Alcock\altaffilmark{3}}

\altaffiltext{1}{Department of Statistics, Harvard University,
1 Oxford Street, Cambridge, MA, 02138 }

\altaffiltext{2}{Harvard-Smithsonian Center for Astrophysics, 60 Garden Street,
Cambridge, MA 02138}

\altaffiltext{3}{Initiative In Innovative Computing, Harvard University,
60 Oxford Street, Cambridge, MA 02138}

\begin{abstract}
We present a new approach to the analysis of time symmetry in  light curves,
such as those in the x-ray at the center of the Scorpius X-1 occultation debate.
Our method uses a new parameterization for such events (the bilogistic event
profile) and provides a clear, physically relevant characterization of each
event's key features. We also demonstrate a Markov Chain Monte Carlo algorithm
to carry out this analysis, including a novel independence chain configuration
for the estimation of each event's location in the light curve. These tools are
applied to the Scorpius X-1 light curves presented in \cite{chang2007},
providing additional evidence based on the time series that the events detected
thus far are most likely not occultations by TNOs.
\end{abstract}


\maketitle 

\section{Introduction}

As new surveys accumulate vast databases of astronomical light curves
\citep{Hodapp2004,Starr2002}, the identification and analysis of events in these
time series is becoming an increasingly important task (e.g. 
\cite{Preston2009,Stewart2009}). One of the topics of such analyses is examining
the symmetry of such events. Traditionally, symmetry has been analyzed by
examining the skewness of each event, sometimes after stacking the light curves
(for examples, see \cite{jones2006mds,jones2008apj}). However, this approach is
not satisfactory for x-ray light curves due to the small number of counts
involved, particularly during suspected occultation events. Additionally, the
nonparametric approach of examining skewness has significantly lower power than
a parametric analysis with a well parameterized model, especially considering
the large mean squared errors of common skewness estimators (see
\cite{joanes1998cms}). This means that, for a given false-positive rate, a
well-designed parametric analysis will have a lower false-negative rate than a
nonparametric approach.

The debate over the nature of the observed Scorpius X-1 events has raged in the
literature since 2006. It began with the publication of \citet{chang2006oxr},
which claimed the detection of at least 58 occultations of Scorpius X-1 by small
(diameter $<$ 100m) TNOs. The size distribution of these small TNOs bears
important clues of the dynamical evolution of the early Solar System (e.g.
\cite{Pan2005,Kenyon2004}), yet the vast majority of TNO population is still way
beyond the limit of direct observation by even the largest ground-based
telescopes. Occultation of background  stars by TNOs is the only present
observational method able to reach the smaller objects
\citep{Bailey1976,Lehner2008,Roques2006,Bickerton2008,biancoetal2009}.  This
finding was challenged later that year by \citet{jones2006mds}. The latter group
claimed that the observed events were unlikely to be occultations by TNOs
because the stacked profile for these events was asymmetric. However, this
analysis was not entirely satisfactory from a statistical perspective; it was
performed on heavily transformed versions of the originally data, did not take
into account the discrete nature of the observations, and provided no measures
of uncertainty.

The response of Chang et al. to this analysis was \citet{chang2007}, wherein
they claimed to have identified 12 events in the Scorpius X-1 data that were
unlikely to have originated from instrumental effects. However, this claim was
challenged again by \citet{jones2008apj}. Jones et al. presented evidence that
the observed events were due to charged particle events in the detectors. Their
argument relied again, in part, on the claim that the observed event were not
temporally symmetric. Unfortunately, their analysis had not proceeded beyond a
simple stacking-based approach (analogous to \cite{jones2006mds}), so the
validity of these conclusions remains uncertain.

Most recently, \cite{liu2008mnras} presented a new analysis of 72-ks of data of
Sco-X1  taken  in the year 2007. \citeauthor{liu2008mnras} concluded that no
significant dips which might be real occultation by $60-100$~m TNOs were
observed.

In this paper we introduce a new parameterization for unimodal events in light
curves, called the bilogistic event profile. This parameterization can describe
a wide range of possible events and naturally yields estimates of physically
interesting quantities such as the full width at half maximum (FWHM) of the
event. Additionally, our parameterization can be applied in both the Poisson
and Gaussian regimes.

Additionally, we demonstrate a computational approach to inference with the
bilogistic event profile using Markov chain Monte Carlo (MCMC) methods. This
includes the introduction of a novel approach to the construction of a proposal
distribution for the center of the event, improving the efficiency of our
simulations.\footnote{In Metropolis-Hastings-type MCMC methods, each step of the
Markov chain consists of two stages. In the first, a new value for the
variable is generated from a proposal distribution with a known density. Then
the proposed value is accepted or rejected as the next value of the chain with a
certain probability (calculated using the target and proposal densities). By
constructing a proposal distribution that is similar to the target distribution
we are attempting to draw from with our MCMC method, we are able to improve the
efficiency of our algorithm.}

Finally, we apply our parameterization and computational approach to occultation
events identified in the Scorpius X-1 RXTE data in \cite{chang2006oxr} and
\cite{chang2007}. Our analysis lends further support to the argument that
the events identified thus far are instrumental in origin, as suggested in
\cite{jones2006mds} and \cite{jones2008apj}.

In \autoref{sec:model}, we describe the details of our data model for the
Scorpius X-1. We present the details of our MCMC approach in
\autoref{sec:compdetails}. The results of applying our method to simulated data
are discussed in \autoref{sec:simresults}. Our primary scientific results on the
Scorpius X-1 light curves comprise \autoref{sec:scox1results}. We conclude in
\autoref{sec:conclusions}.

\section{Model}
\label{sec:model}

\subsection{Context \& Specification}
We have a time series of counts $\{y_t\}$, $t \in \{ 0, \cdots, T-1 \}$. We
assume that each count is distributed as an independent Poisson random variable,
conditional on the source intensity at time $t$:
\begin{align*}
 y_t \sim Poisson(\lambda_t).
\end{align*}
\noindent We also believe that an event of some kind has occurred in this time
series. For the sake of this discussion, we assume that this event is
characterized by a dimming of the source (e.g., a transit). However, this
framework can be extended trivially to events character by increased intensity
(e.g., supernovae) or the high-count regime (with a Gaussian distribution
replacing the Poisson). The key constraint is that we know \textit{a priori} the
sign of the expected deviation from the baseline intensity. We characterize the
intensity $\lambda_t$ as
\begin{align*}
 \lambda_t &= c - \alpha \, g(t;\tau,\overrightarrow{\theta})
\end{align*}
where $\lim_{t \rightarrow \infty} g(t;\tau,\overrightarrow{\theta}) = \lim_{t
\rightarrow -\infty} g(t;\tau,\overrightarrow{\theta}) = 0$ and $\sup_\mathbb{R}
g(t;\tau,\overrightarrow{\theta}) = g(\tau;\tau,\overrightarrow{\theta}) = 1$.
We will call $g(t;\tau,\overrightarrow{\theta})$ the event profile, as it
characterizes the pattern of intensity changes throughout the event. The
parameter $\tau$ is the time index about which the event is hypothesized to be
time-symmetric, and $\overrightarrow{\theta}$ is a vector of parameters
characterizing the event. $c$ sets the baseline intensity of the source, and
$\alpha$ sets the peak size of the deviation from the baseline intensity during
the event. All of these parameters must be estimated from the data.

\noindent  Putting all of these points together with the log-likelihood for the
model, we obtain:
\begin{align}
 \ell(c,\alpha,\tau,\overrightarrow{\theta};\overrightarrow{y})
  &= \sum_{t=0}^{T-1} [y_t \ln{(\lambda_t)} - \lambda_t] + \rm{const} 
  \end{align}
  
 \noindent Following Bayesian approach we can express the posterior probability 
 as a function of the likelihood and the prior probabilities as: 
 
  \begin{align}
 p(c,\alpha,\tau,\overrightarrow{\theta} \mid \overrightarrow{y}) &\propto
  \ell(c,\alpha,\tau,\overrightarrow{\theta};\overrightarrow{y}) \times
  p(c,\alpha,\tau,\overrightarrow{\theta}) \label{eq:posteriordef}
\end{align}
\noindent Having obtained the posterior of our parameters (up to a normalizing
constant), we have all of the information needed to make inferences about the
parameters of interest. We can characterize our inferences about the locations
of parameters via their posterior means and medians. Our uncertainty about these
parameters can be characterized by $68\%$ posterior intervals or posterior
standard deviations. However, it still remains to specify a functional form for
the event profile $g(t;\tau,\overrightarrow{\theta})$.

\subsection{Bilogistic event profile}
The choice of event profile $g(t;\tau,\overrightarrow{\theta})$ is quite
important to this analysis. One must achieve the proper balance between
parsimony in parameterization and giving the model enough flexibility to fit a
wide range of event shapes. To that end, we introduce the bilogistic event
profile:

\begin{align}
 g(t;\tau,h_1,h_2,k_1,k_2) = \frac{1 + e^{\frac{-h_t}{k_t}}}
 {1 + e^{\frac{|t-\tau|-h_t}{k_t}}}\\
 h_t = \left\{\begin{array}{l l}
                     h_1 & t < \tau\\
                     h_2 & t \geq \tau
                    \end{array}\right.\\
 k_t = \left\{\begin{array}{l l}
                     k_1 & t < \tau\\
                     k_2 & t \geq \tau
                    \end{array}\right.
\end{align}

This event profile is characterized by four parameters (in addition to the
location parameter $\tau$). The first two parameters ($h_1$ and $h_2$)
characterize the (approximate) half-maxima of the event's amplitude, represented
as deviations from the event's center; we will refer to them as the
``half-life'' parameters. This holds only approximately because of the
correction for continuity in the numerator (the $e^{\frac{-h_t}{k_t}}$ term),
which ensures that the event profile is continuous at time $\tau$. Thus,
$(h_1-h_2)$ characterizes the departure of our event from symmetry in terms of
the half-maxima, and $(h_1+h_2)$ characterize the (approximate) full width at
half maximum for our event.

\begin{figure}
 \epsscale{1}
 \plotone{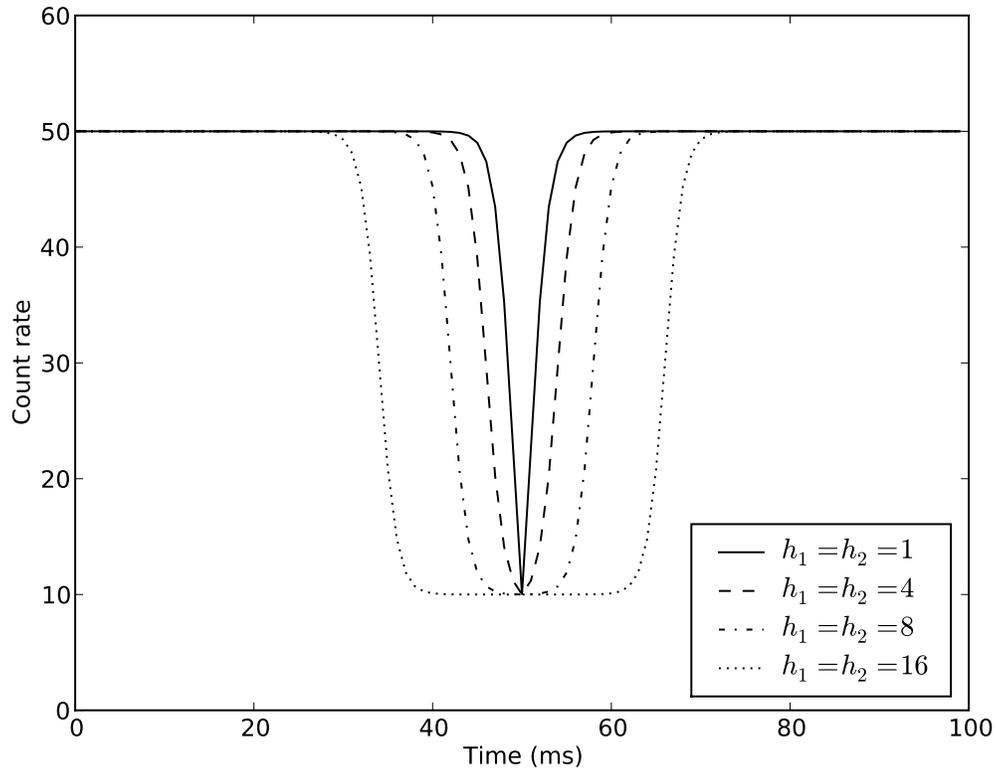}
 \caption{Bilogistic event profile for $c=50$, $\alpha=40$, $k_1=k_2=1$, and
 $\tau=50$. Varying $h_1$ and $h_2$ from $1$ to $16$. \label{fixk}}
\end{figure}

\begin{figure}
 \epsscale{1}
 \plotone{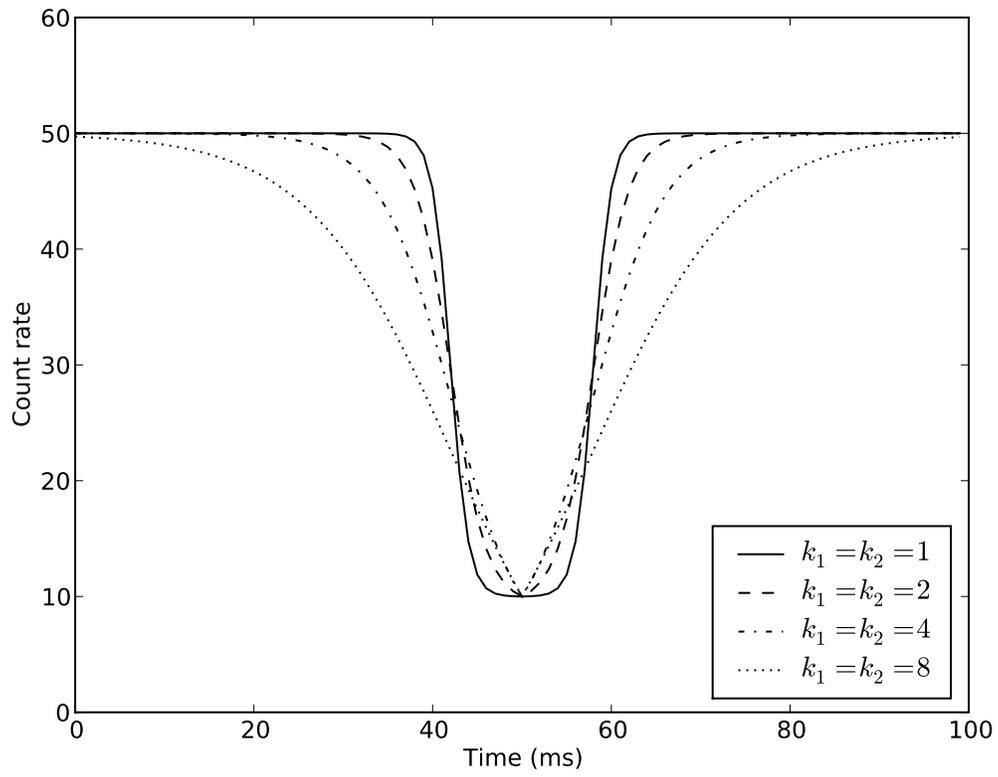}
 \caption{Bilogistic event profile for $c=50$, $\alpha=40$, $h_1=h_2=8$, and
 $\tau=50$. Varying $k_1$ and $k_2$ from $1$ to $8$. \label{fixh}}
\end{figure}

The second set of parameters ($k_1$ and $k_2$) characterize the rate at which
the intensity changes during our event; we will refer to them as the
``curvature'' parameters. The first ($k_1$) characterizes how rapidly our
source's intensity diminishes at the beginning of our event, and the second
characterizes how rapidly the intensity returns to the baseline at the end of
our event. If $k_1 << h_1$ and $k_2 << h_2$, as is often the case, the curvature
parameters can been viewed as pivoting our event profile through the
half-intensity points set by $h_1$ and $h_2$, with higher values corresponding
to sharper changes in intensity. These features of the bilogistic event profile
can be seen in Figures \hyperref[fixk]{\ref*{fixk}} and
\hyperref[fixh]{\ref*{fixh}}, where we fix one set of parameters ($h_1$ and
$h_2$ or $k_1$ and $k_2$) and vary the other.

\phantomsection
\label{priordiscussion}
To perform inference using this model, we first consider it in a Bayesian light,
assuming flat priors on all parameters except $\tau$. We place one restriction
on $\tau$ a priori.
We assume that the event is centered within the interior of our light curve. In
this context, we take interior to mean far enough from the boundaries that the
event is contained within our light curve. This assumption is reasonable for our
analysis because the light curves we are working with have been extracted and
preprocessed to contain one complete event each. Thus, we impose a prior
probability of zero on values of $\tau$ within $m$ milliseconds of the edge of
the light curve for our analysis; for the analysis presented here, we set $m =
10$ milliseconds. This assumption also improves the computational properties of
our posterior simulations by preventing them from becoming ``stuck'' at values
of the $\tau$ parameter that are near the ends of the light curve. The end
result of these restrictions is that our prior on $\tau$ is uniform on the set
$[m,T-m]$.\footnote{However, for  other analyses of time symmetry in events
(such as the analysis of longer duration events that are not fully contained in
a light curve), these assumptions could be relaxed.}

We can then use a Markov chain Monte Carlo (MCMC) technique (the
Metropolis-Hastings algorithm) to draw from the posterior distribution of our
parameters. From these draws, we can calculate quantities such as the posterior
median and 68\% posterior intervals for $(h_1-h_2)$ and $(k_1-k_2)$ to
characterize the deviation of our events from symmetry and our uncertainty about
these properties. For an excellent introduction to the use of the
Metropolis-Hastings algorithm for statistical inference, please consult
\cite{chib1995umh}.

\section{Computational details of posterior simulations}
\label{sec:compdetails}

Although the model presented above is quite appealing, the resulting posterior
distribution for our parameters of interest is not analytically tractable. Thus,
to make inferences about these parameters, we simulate a series of samples from
their joint posterior distribution. We use a blocked Metropolis-Hastings
algorithm for these simulations. This means that, instead of drawing new values
for all seven of our parameters at once and accepting or rejecting them based on
the Metropolis-Hastings rule, we vary only a subset of the parameters at time,
holding all others fixed. The blocks used are: (1) $\tau$, (2) $c$ and $\alpha$,
and (3) $h_1$, $h_2$, $k_1$ and $k_2$. So, for each iteration of our algorithm,
we first draw a new value of $\tau$ from our proposal distribution and choose
whether to accept it based on the Metropolis-Hastings acceptance probability.
This acceptance probability is calculated holding the values of all other
parameters ($c$, $\alpha$, $h_1$, $h_2$, etc.) fixed. Next, we draw new values
of $c$ and $\alpha$ from our proposal distribution for these parameters and
calculate an analogous acceptance probability. Finally, we perform a similar
procedure on the shape parameters ($h_1,h_2,k_1$, and $k_2$). Blocking in this
way allows us to take advantage of the structure in the model in constructing
our simulation algorithm, improving its efficiency. For more information on
blocked Metropolis-Hastings methods, please consult \S 6.2 of
\cite{chib1995umh}.

\subsection{Initialization}

We look to the data to obtain starting values for our MCMC algorithm. The first
step is to obtain a smoothed version of our signal $y_t$, which we will denote
$\tilde{y}_t$. This is generated by taking a centered moving average of the
original signal $y_t$. For the analyses presented here, a window ten
milliseconds wide was used, although this can be modified. The important point
on the selection of the window is that it should be narrow enough to localize
the event while still performing some noise reduction. If the window is too
wide, our smoothed light curve will not capture the structure of the
the event; as a result, it would not be useful in determining initial values for
the magnitude ($\alpha$) or in determing the location of the event in the
light curve. However, if the window is too narrow, it will not smooth the
light curve well. As a result, our initializations would be far noisier than
necessary.

Using $\tilde{y}_t$, we can construct our initialization values and a proposal
distribution for $\tau$. Focusing on initialization first, we set $c_0 =
\operatorname{median}{(\tilde{y}_t)}$ and $\alpha_0 = c_0 - \min_{t}(y_t)$. We
use a robust measure of typical baseline rate (the median). The motivation for
this initialization of c is that, if the event is brief relative to the duration
of the light curve and our smoothing window is not too long, our smoothed light
curve should be near the baseline count rate $c$ for most times. In contrast, if
we simply used the overall mean, it would be contaminated by the event of
interest. Our initialization for $\alpha$ is relatively straightforward; we take
our estimate of the baseline rate ($c_0$) and subtract the minimum observed
count rate to obtain an estimate of the magnitude of our event.

To initialize $h_1$ and $h_2$, we take a similar approach. First, we assume that
a symmetric initialization will be an acceptable starting point for our
analysis. We then calculate $h_0 = \frac{1}{2} \# \left\{ t : y_t < \frac{
\operatorname{median}{(\tilde{y}_t)} + \min_{t} (y_t)}{2} \right\}$ and set
$h_1$ and $h_2$ to $h_0$.\footnote{$\#$ is the cardinality operator, which gives
the number of items in the given set} Based on the above discussion of $c_0$
and $\alpha_0$, this approach to initializing $h_1$ and $h_2$ appears quite
natural. We are using our initial estimates of $c$ and $\alpha$ to estimate the
FWHM of our event and attributing equal parts of this time to $h_1$ and $h_2$.
Finally, we initialize $k_1$ and $k_2$ manually. Our posterior simulations were
quite insensitive to the initializations of these parameters; values between 1
and 20 yielded similar results.

\subsection{Metropolis-Hastings steps}

Keeping in mind the prior discussed in \hyperref[priordiscussion]{\S
\ref*{priordiscussion}}, we can obtain a proposal distribution for $\tau$ using
the procedure outlined in Algorithm 1.

\begin{algorithm}
\label{tauproposal}
\caption{Construction of proposal distribution for $\tau$}
\begin{algorithmic}
  \REQUIRE $\tilde{y}_t$ for $t \in \{1,\ldots,T\}$, $m \in \mathbb{N}$,
   $p > 0$, $\epsilon > 0$
  \STATE $\tilde{w}_t \leftarrow (\max_{t}{(\tilde{y}_t)} - \tilde{y}_t)^p$
  \STATE $\tilde{w}_t \leftarrow \tilde{w}_t + \epsilon$
  \IF{$t < m$ or $(T-t) < m$}
    \STATE $\tilde{w}_t \leftarrow  0$
  \ENDIF
  \STATE $w_t \leftarrow \frac{\tilde{w}_t}{\sum_{t} \tilde{w}_t}$
\vspace{0.2cm}
  \STATE Calculate points along cumulative distribution function (CDF) as $p_t =
  \sum_{s \le t} w_s$
  \STATE Calculate CDF $F(t)$ via linear interpolation of $\{p_t\}$
  \STATE Calculate quantile function $q(p)$ via linear interpolation
  \STATE Calculate density function $f(t)$ as derivative of $F(t)$ for
  $t \notin \mathbb{N}$
\end{algorithmic}
\end{algorithm}

\noindent At the conclusion of this procedure, we have the density, cumulative
density, and quantile functions for our proposal distribution for $\tau$.
Through a series of numerical experiments, we found that values of $p$ between
two and five typically produced the best proposal distributions; $p$ was set to
four in the subsequent analyses. The $\epsilon$ term serves to ensure that all
$w_t$ are non-zero, avoiding an issue with the interpolation of our quantile
function (where a $w_t$ of zero leads to a vertical segment). We use linear
interpolation because it is the only spline method that produces identical
functions when the axes are exchanged. This is vital in our application because
we are conducting seperate interpolations of $\{p_t\}$ to obtain our CDF and its
inverse (the quantile function). If we had used any other type of polynomial
interpolation, the inverse of the calculated spline for the CDF would not be
another polynomial spline. However, with a linear spline, the inverse of the
interpolated function is simply another linear spline. Thus, we can avoid
complex calculations and use standard interpolation routines to calculate our
quantile function. Our method of generating a proposal for our location
parameter is quite versatile, adapting naturally to the presence of multiple
potential events in the light curve. We will discuss some potential applications
of it beyond this analysis in \autoref{sec:conclusions}.

The proposal generated by the above procedure allows us to use an independence
chain Metropolis-Hastings method for $\tau$. With this class of methods, the
proposal distribution for the parameter being simulated is independent of that
parameter's current value in the MCMC simulation. If the proposal is
well-constructed, this can greatly improve the convergence properties of the
simulations; in the ideal case, we would make our proposal the same as our
posterior to maximize efficiency. However, for such a procedure to be effective,
one must have some prior knowledge of the parameters's posterior distribution.
That is what the above procedure extracts from the light curve.


Using the initial values and proposal distribution obtained above, we construct
proposal distributions for the parameters of our model (excepting $\tau$). All
proposals are Gaussian and are thus characterized by their means and standard
deviations:
\begin{align}
  (c, \alpha) \: :& \; \mu = (c_0, \alpha_0), \;
   \sigma = (\sqrt{\frac{c_0}{T}}, \sqrt{\alpha_0})\\
  (h_1, h_2) \: :& \; \mu = (h_{1}^{i-1}, h_{2}^{i-1}), \;
   \sigma = (\frac{\sqrt{h_0}}{2}, \frac{\sqrt{h_0}}{2})\\
  (k_1, k_2) \: :& \; \mu = (k_{1}^{i-1}, k_{2}^{i-1}), \;
   \sigma = (\frac{\sqrt{k_0}}{2}, \frac{\sqrt{k_0}}{2})
\end{align}

With these proposals and the blocking scheme described previously, we have fully
specified our MCMC simulation approach. We now demonstrate it's effectiveness
with simulated data.

\section{Results on simulated symmetric events}
\label{sec:simresults}

\begin{figure}
  \epsscale{1}
  \plotone{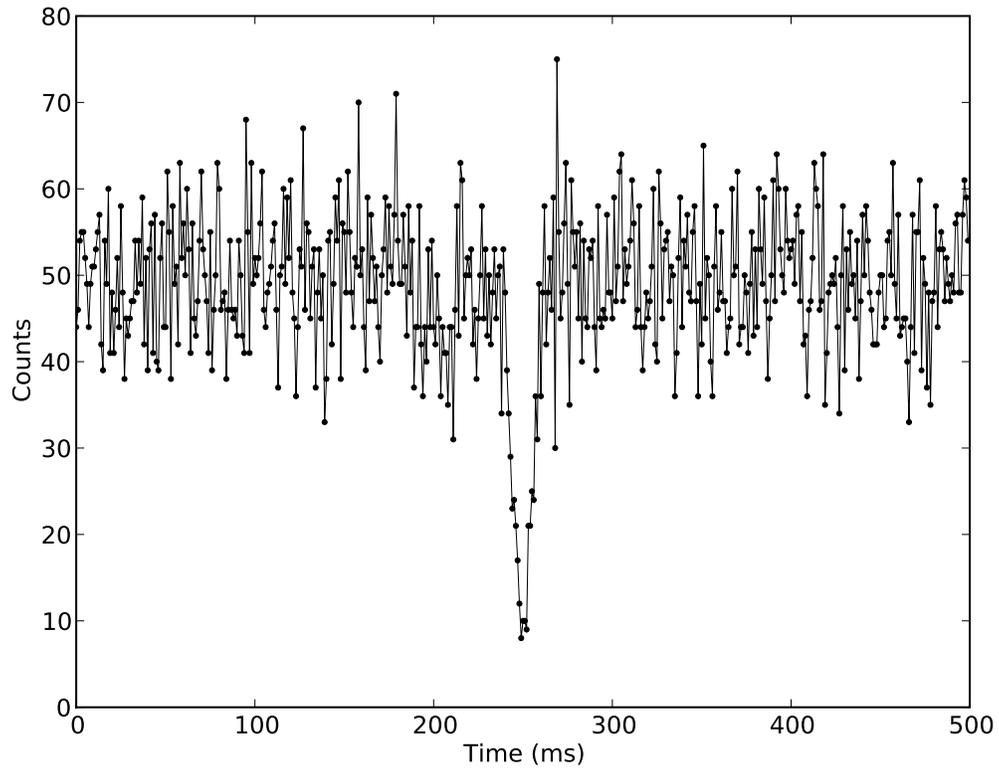}
  \caption{Simulated light curve for symmetric event
  \label{simlightcurve}}
\end{figure}

To test our approach, we simulated a 500 time steps light curve with a symmetric
event resembling those observed in the Scorpius X-1 data. For this simulation,
we set $c = 50$, $\alpha = 40$, $h_1 = h_2 = 5$, $k_1 = k_2 = 2$, and $\tau =
300$. The resulting light curve can be seen in \autoref{simlightcurve}.

\begin{figure}
  \epsscale{1}
  \plotone{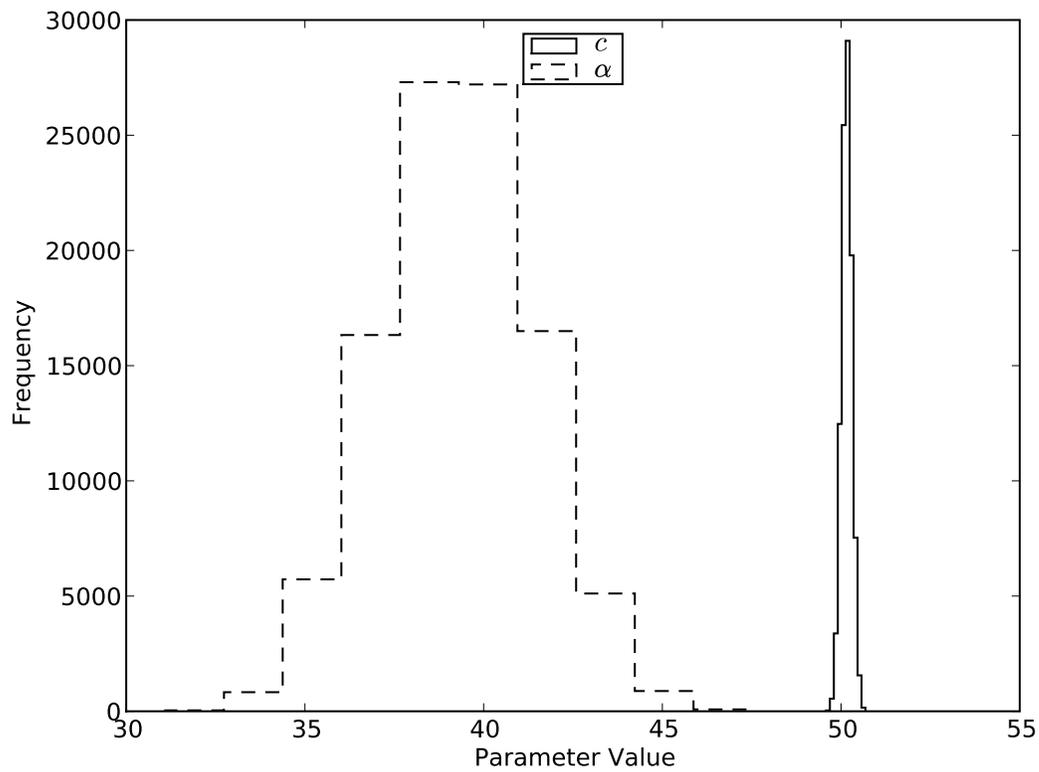}
  \caption{Histogram of posterior distributions of $c$ and $\alpha$ for
   simulated event \label{simmagnitudehist}}
\end{figure}

\begin{figure}
  \epsscale{1}
  \plotone{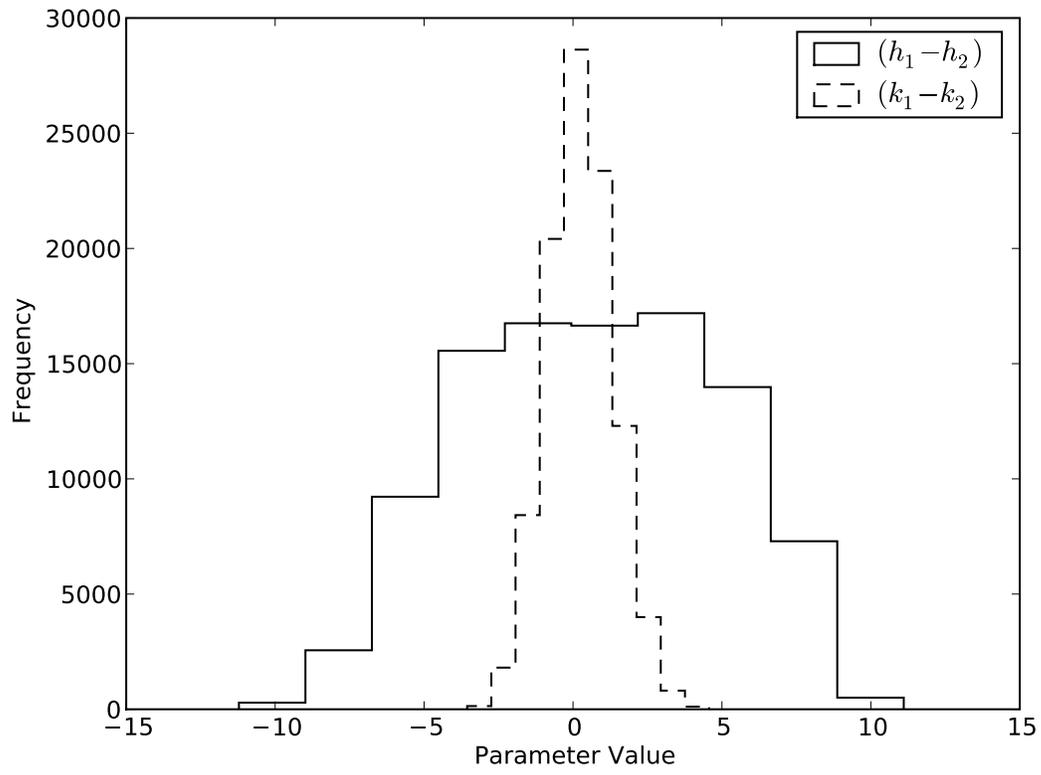}
  \caption{Histogram of key symmetry-related parameters for simulated event
  \label{simsymmetryhist}}
\end{figure}

\begin{figure}
  \epsscale{1}
  \plotone{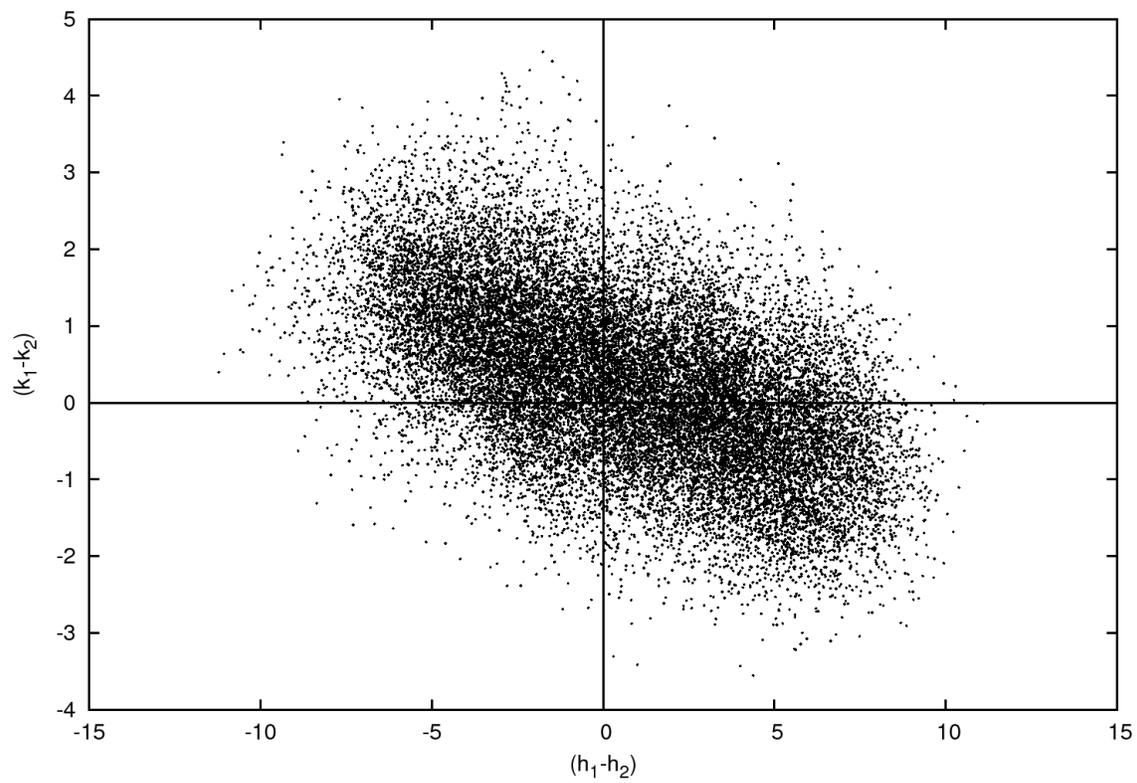}
  \caption{$(k_1-k_2)$ vs. $(h_1-h_2)$ for simulated event
   \label{simsymmetryscatter}}
\end{figure}

The results of the posterior simulations are summarized in
\autoref{simmagnitudehist}, \autoref{simsymmetryhist}, and
\autoref{simsymmetryscatter}. From the first of these
(\autoref{simmagnitudehist}), we observe that the posterior distributions for
both the baseline count rate ($c$) and the magnitude of the event ($\alpha$) are
centered around their true values of 50 and 40, respectively. The posterior on
$c$ is significantly more concentrated than the posterior on $\alpha$, as we
would expect; we can use nearly every point in the light curve to infer the
value of $c$, but we can only use the points during the event to infer the value
of $\alpha$. The posterior distributions of $(h_1-h_2)$ and $(k_1-k_2)$ are
considerably more diffuse, as \autoref{simsymmetryhist} demonstrates. These
posteriors are, however, centered near the true value of 0. The relationship
between the posterior distributions of these two quantities can be seen from
\autoref{simsymmetryscatter}. There appears to be a significant negative
dependence between the two deviations.

The properties of our estimates for symmetry-related parameters ($(h_1-h_2)$ and
$(k_1-k_2)$) are further supported by a second set of simulations. Using the
parameter values given above, we simulated 1000 light curves and ran our
algorithm on each one. This allows us to quantify some frequentist properties of
our Bayesian estimators, such as the coverage of our posterior
intervals.\footnote{The coverage of an interval estimator is defined as the
percentage of intervals that will cover the true value of the parameter of
interest across repeated sampling. By checking the coverage of our key
estimators, we can quantify their classical error rates in addition to their
Bayesian properties. For further details, \cite{wasserman2004asc} provides an
excellent reference in \S 6.3.2 and \S 11.9.}  The results of these simulations
are summarized in \autoref{multiplesimsymmetryhist}. Both distributions appear
centered around zero. Coverage properties are not perfectly calibrated (as this
is a Bayesian method), but they appear reasonable. 938 of our 1000 $68\%$
posterior intervals for $(h_1-h_2)$ included zero, as did 816 of our $68\%$
posterior intervals for $(k_1-k_2)$.

Overall, the results of this exercise with simulated events serve to increase
our confidence in the validity of our method. We now apply our model and
algorithm to the analysis of the Scorpius X-1 dip events.

\section{Results on Scorpius X-1 events}
\label{sec:scox1results}

\begin{figure}
 \epsscale{1}
\begin{minipage}[t]{0.95\textwidth}
\begin{center}
 \includegraphics[width=3in,height=2in]{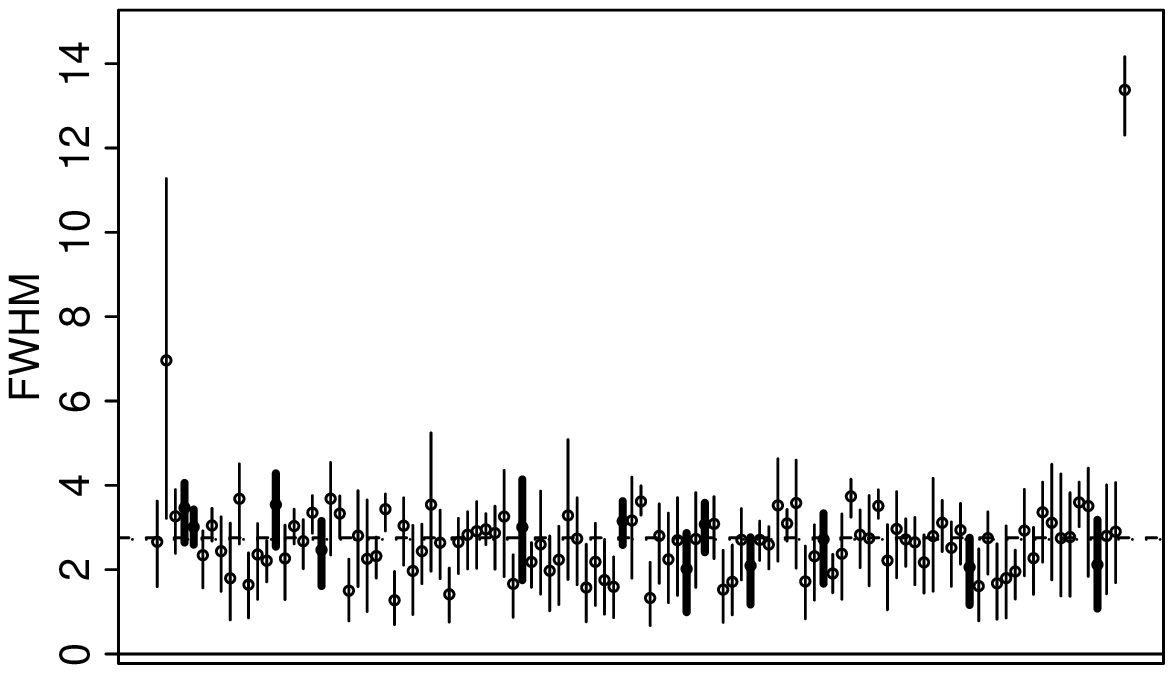}\\
  \vspace{-1cm}
 \includegraphics[width=3in]{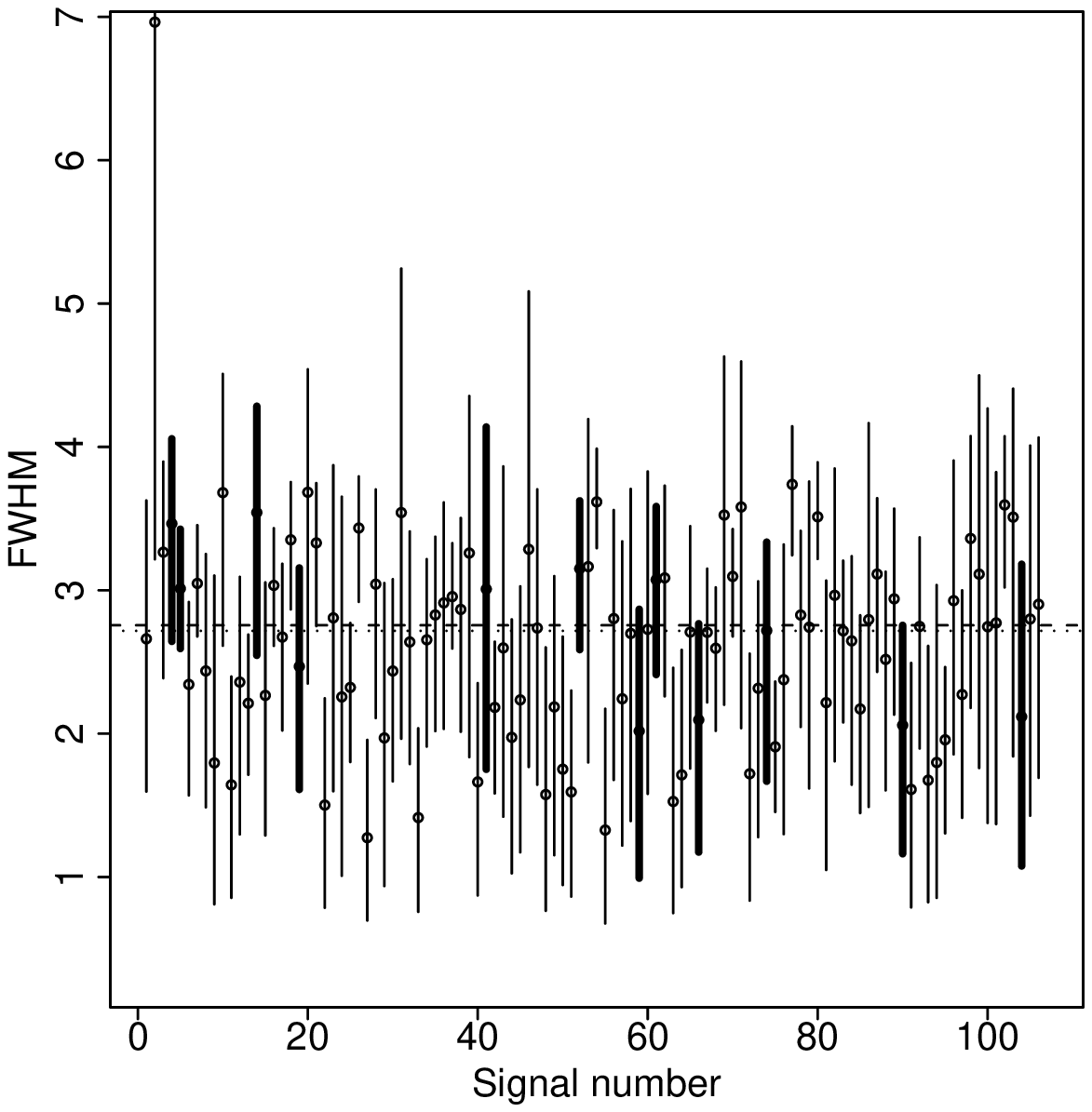}
  \caption{68\% posterior Intervals \& medians for FWHM $(h_1+h_2)$ for
\citet{chang2007} occultation events.\label{fihmintervals2007} Second plot
is zoomed.}
\end{center}
\end{minipage}
\end{figure}

\begin{figure}
 \epsscale{1}
\begin{minipage}[t]{0.95\textwidth}
\begin{center}
 \includegraphics[width=3in,height=2in]{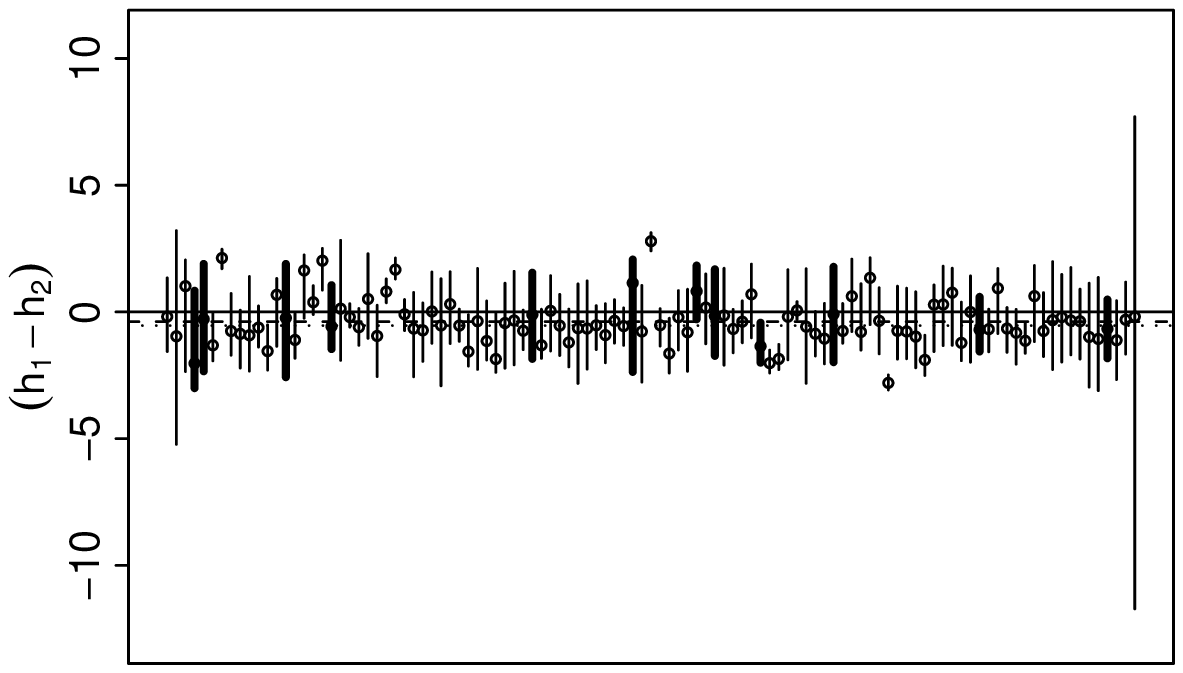}\\
 \vspace{-1cm}
 \includegraphics[width=3in]{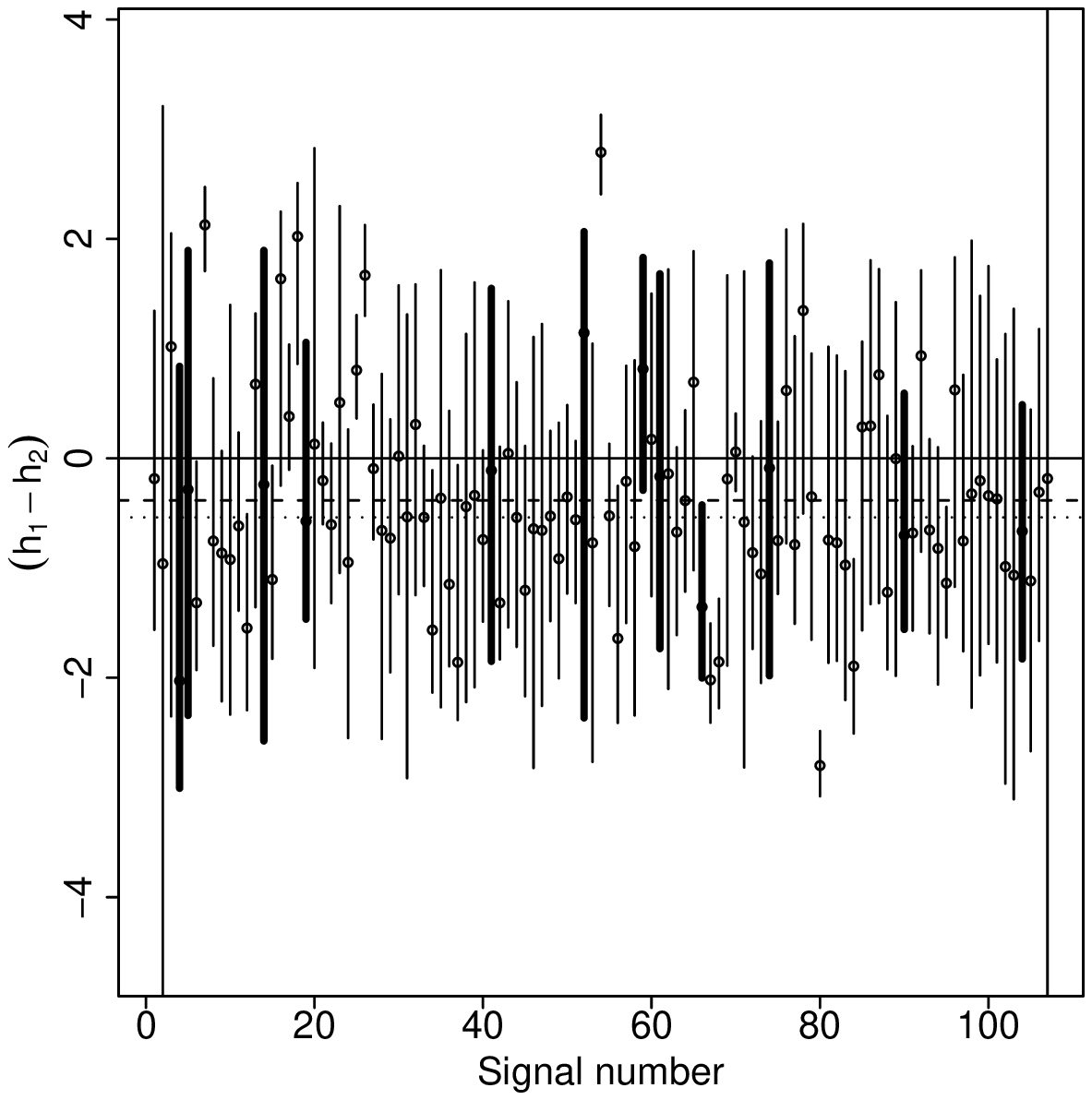}
 \caption{68\% posterior Intervals \& medians for $(h_1-h_2)$ for
\citet{chang2007} occultation events.\label{hdiffintervals2007} Second plot
is zoomed.}
\end{center}
\end{minipage}
\end{figure}

\begin{figure}
 \epsscale{1}
  \begin{minipage}[t]{0.95\textwidth}
\begin{center}
 \includegraphics[width=3in,height=2in,clip=true,trim=0 1cm 0 0]{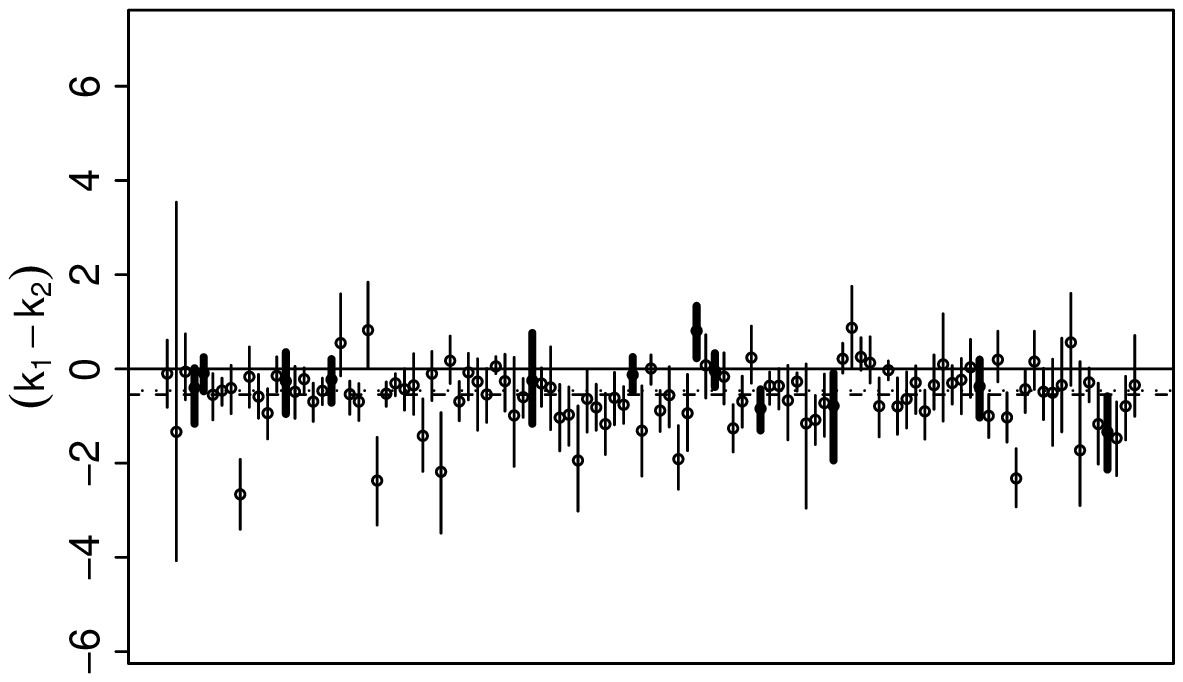} \\
 \vspace{-0.3cm}
 \includegraphics[width=3in]{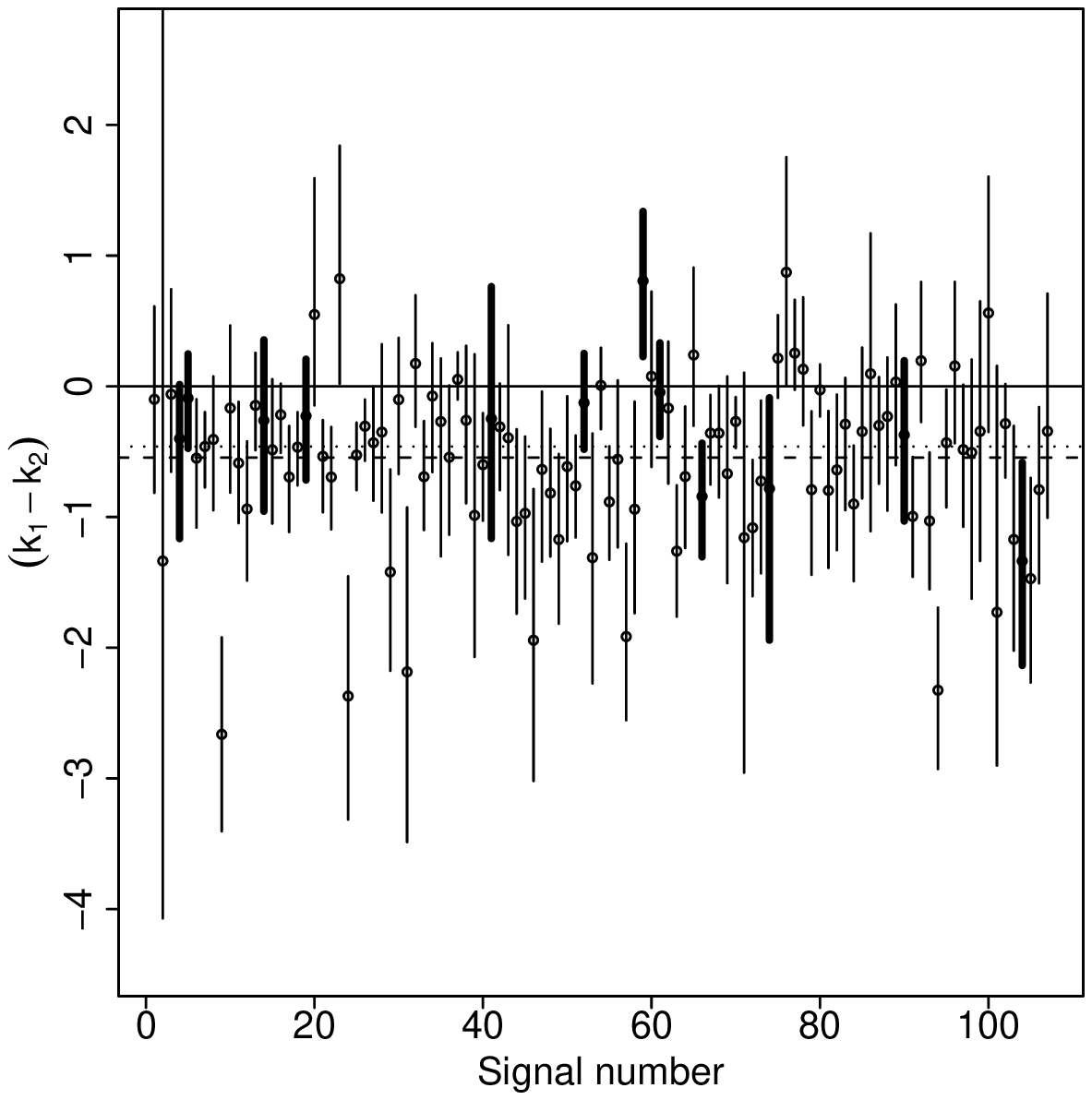}
 \caption{68\% posterior Intervals \& medians for $(k_1-k_2)$ for
\citet{chang2007} occultation events.\label{kdiffintervals2007} Second plot
is zoomed.}
\end{center}
\end{minipage}
\end{figure}

Using the above model, we analyzed all 107 events from \citet{chang2007}. For
each event, we ran our MCMC algorithm for $400,000$ iterations, discarding the
first $200,000$ as burn-in\footnote{In MCMC simulations, it is common to discard
the earliest portion of the results as burn-in to ensure that we are only basing
our conclusions on draws from the Markov chain's stationary distribution (which
is, by design, our posterior).}. Run in parallel on Harvard's Odyssey cluster,
these runs required approximately 400 seconds in total (approximately $43,000$
CPU seconds). Typical acceptance rates for the individual Metropolis-Hastings
steps were between 10\% and 40\%. The results of these simulations are
summarized in \autoref{fihmintervals2007}, \autoref{hdiffintervals2007}, and
\autoref{kdiffintervals2007}.

\begin{figure}
  \epsscale{1}
  \plotone{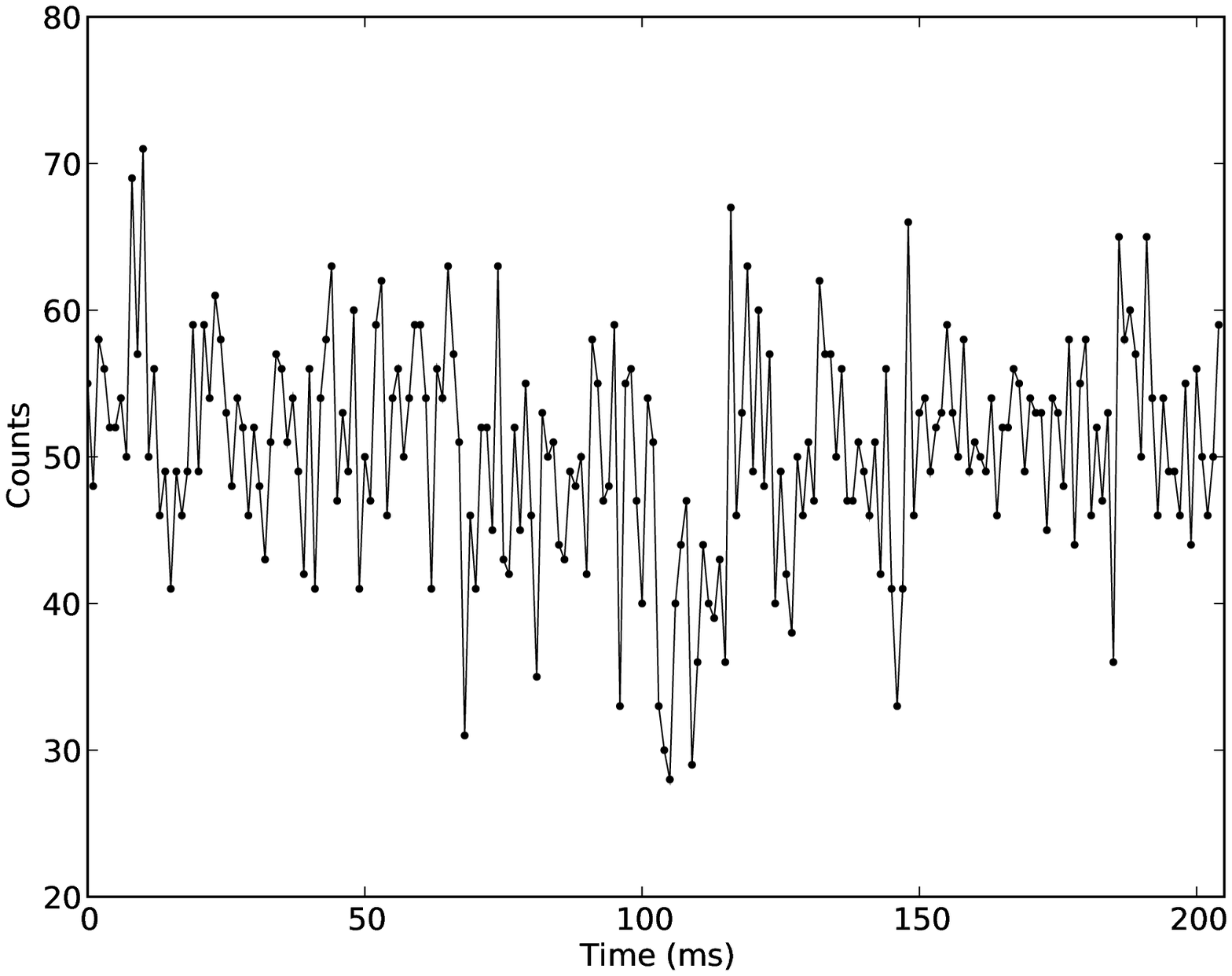}
  \caption{Light curve for event 2 from \citet{chang2007}.
   \label{event002chang2007}}
\end{figure}

\begin{figure}
  \epsscale{1}
  \plotone{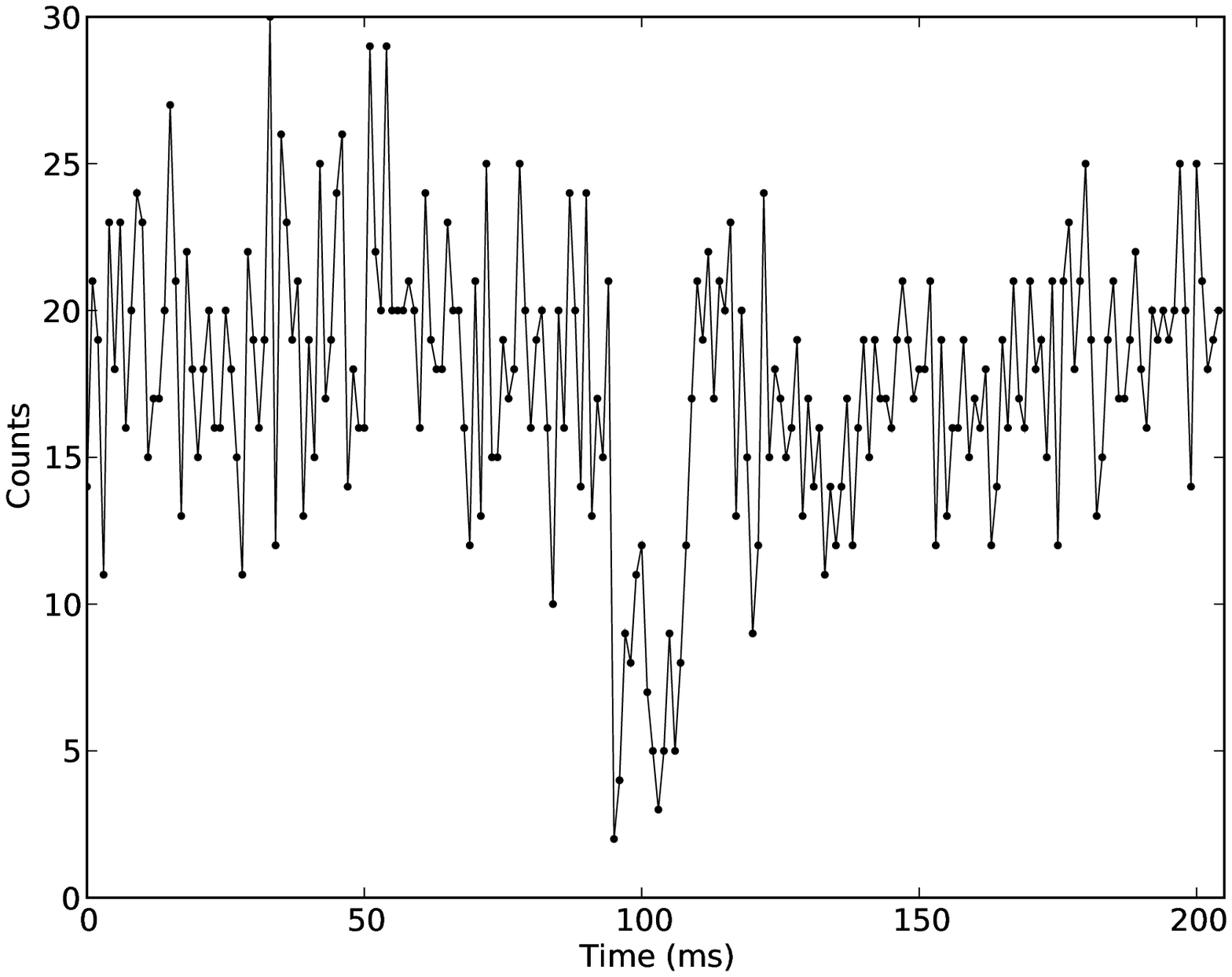}
  \caption{Light curve for event 107 from \citet{chang2007}.
   \label{event107chang2007}}
\end{figure}

From \autoref{fihmintervals2007}, we see that the FWHM of a typical event from
\citet{chang2007} is between 2 and 4 milliseconds (with $68\%$ posterior
probability). Events 2 and 107 appears to be the only significant deviation from
this pattern with FWHMs of approximately 7 and 13 milliseconds, respectively.
Visual inspection of these events (in Figures
\hyperref[event002chang2007]{\ref*{event002chang2007}} and
\hyperref[event107chang2007]{\ref*{event107chang2007}}) reveals that both are
unusual in appearance. Event 2 is characterized by a more gradual change in
intensity than most and appears to have two minimia. Event 107 appears to have
two clear, significant minima.

\begin{figure}
  \epsscale{1}
  \plotone{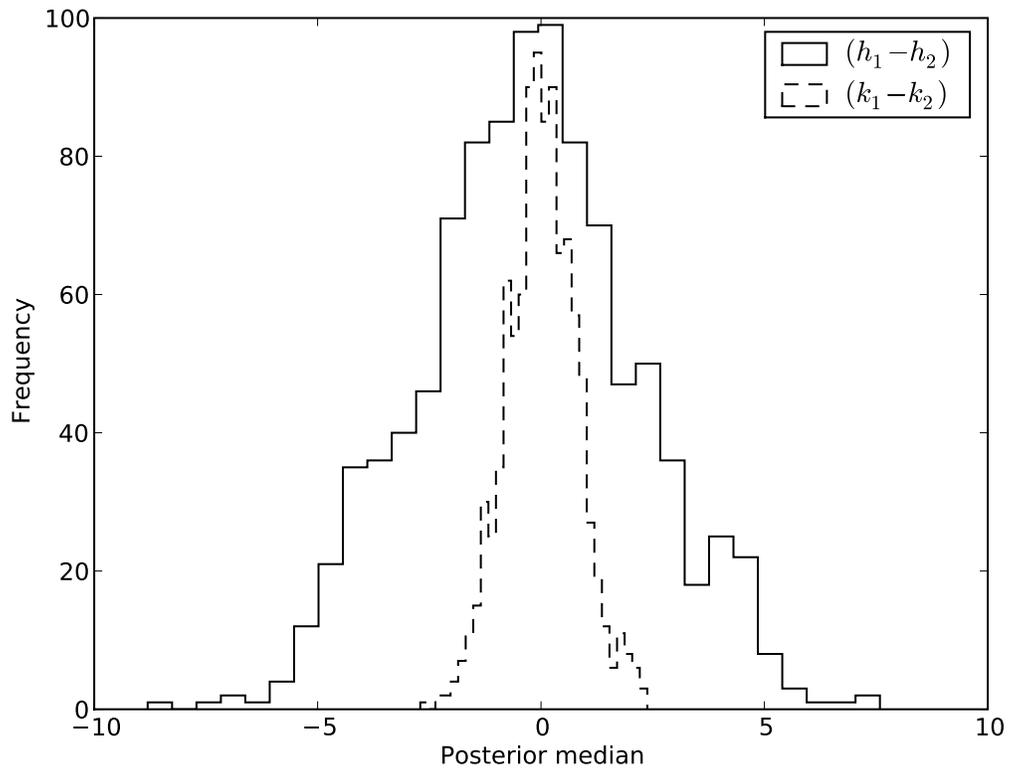}
  \caption{Distributions of posterior medians of $(k_1-k_2)$ and
   $(h_1-h_2)$ for 1000 simulated events
   \label{multiplesimsymmetryhist}}
\end{figure}

\begin{figure}
 \epsscale{1}
 \plotone{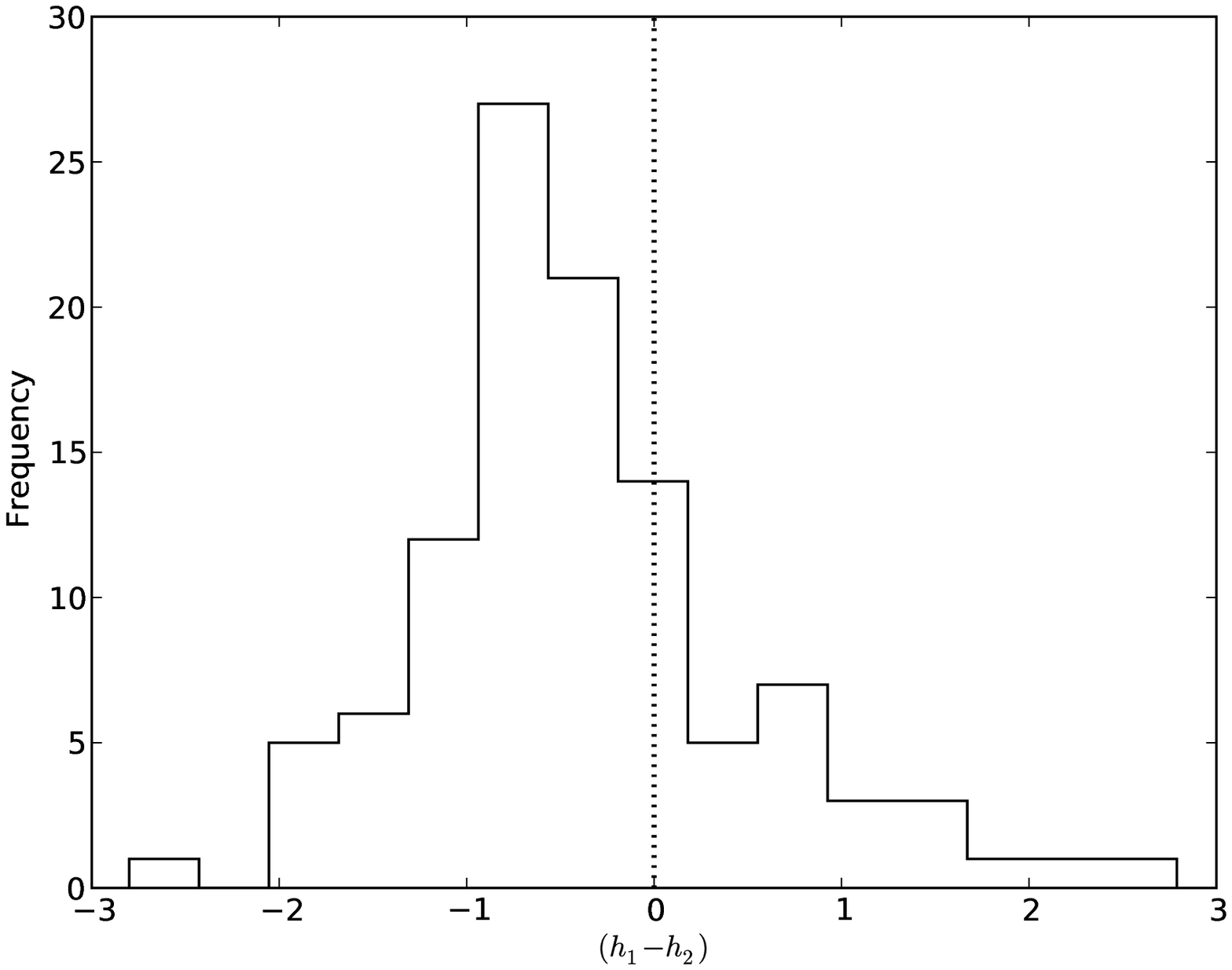}
 \caption{Histogram of posterior medians for $(h_1-h_2)$ for
\citet{chang2007} occultation events.\label{hdiffhist2007}}
\end{figure}

The posterior intervals for $(h_1-h_2)$ in \autoref{hdiffintervals2007} appear
to be relatively evenly scattered around zero. However, although most of these
intervals (90 of 107) cover zero, there is notable tendency towards negative
values of $(h_1-h_2)$. This can be seen in the posterior median, with 81 of the
107 posterior medians below zero. Using a classical sign test of the hypothesis
that each posterior median is equally likely to be positive or negative
\citep{signtest1946}, we obtain a p-value of $9.4 \times 10^{-8}$, indicating
that there is a significant deviation from symmetry in the relative half-maxima
for this collection of events. The sign test is conducted by first counting the
number of positive posterior medians. We then calculated the probability of
observing a result at least as far from an even split of positive and negative
medians as was observed, assuming that each median is equally likely to fall
above or below zero. In this case, the p-value was calculated as $2 \times P(X
\le 26 | p = \frac{1}{2})$, where $X$ is the number of observed positive
medians. Under the null hypothesis that positives and negatives are equally
likely, $X \sim \mbox{Bin}(107,\frac{1}{2})$, so the preceding probability can
be easily calculated using standard techniques for the binomial distribution.
The pattern of negative deviations in $(k_1-k_2)$ can also be seen in
\autoref{hdiffhist2007}; the deviation of this distribution from the symmetric
case (seen in \autoref{multiplesimsymmetryhist}) is clear. Events 2 and 107
appear somewhat anomalous in \autoref{hdiffintervals2007}; this is consistent
with the unusual structure of these events discussed above.

\begin{figure}
 \epsscale{1}
 \plotone{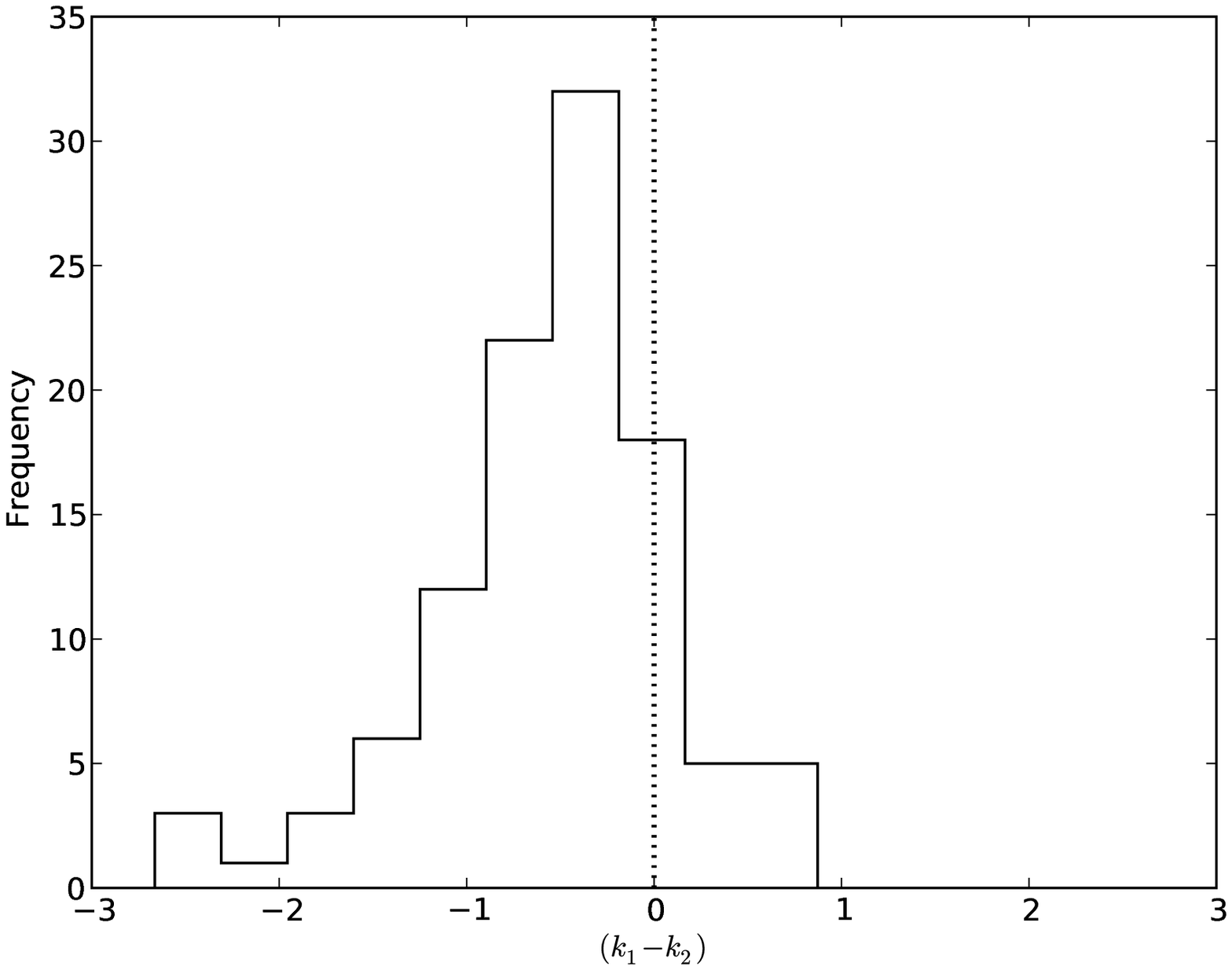}
 \caption{Histogram of posterior medians for $(k_1-k_2)$ for
\citet{chang2007} occultation events.\label{kdiffhist2007}}
\end{figure}

Finally, the posterior medians for $(k_1-k_2)$ are concentrated below zero (90
of 107 are negative). Inspection of the posterior intervals in
\autoref{kdiffintervals2007} indicates that very little posterior support is
typically found above zero (only 55 of the 107 intervals included zero). A
classical sign test on the posterior medians (as above) yields a p-value of $3.5
\times 10^{-13}$. This indicates that the observed pattern of posterior medians
is quite unlikely to occur if the true distribution of posterior medians is
evenly distributed above and below zero. This is consistent with the findings of
\cite{liu2008mnras}, as a negative value of $(k_1-k_2)$ indicates that the count
rate decreased more quickly than it increased. The pattern of negative
deviations in $(k_1-k_2)$ can also be seen in \autoref{kdiffhist2007}; the
deviation of this distribution from the symmetric case (seen in
\autoref{multiplesimsymmetryhist}) is again apparent.

\begin{figure}
 \epsscale{0.9}
 \plotone{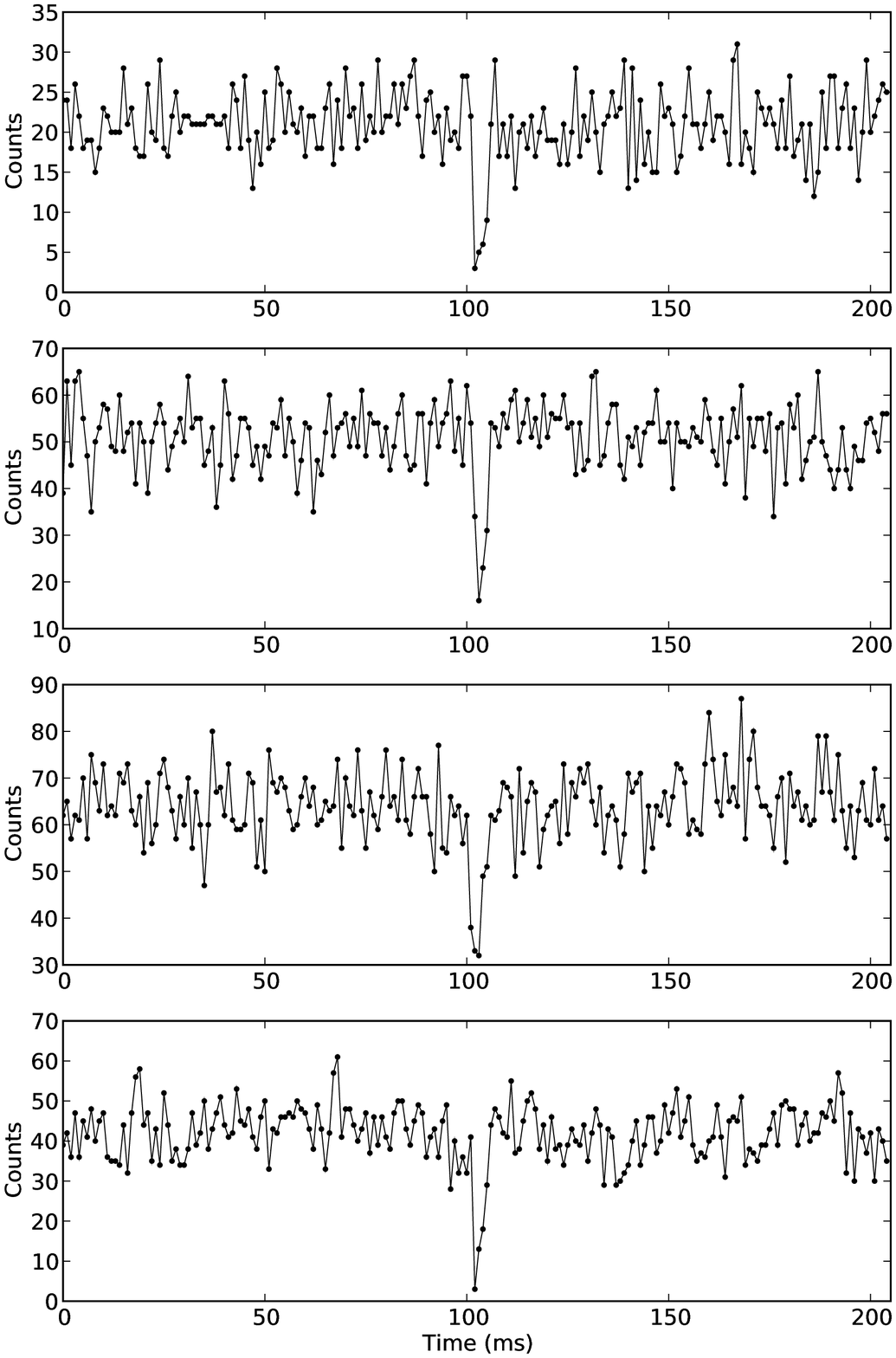}
 \caption{Events 4, 19, 41, and 66 from \cite{chang2007}
 \label{typicalevents}}
\end{figure}

The concentration of $(h_1-h_2)$ and $(k_1-k_2)$ across events indicates
that the observed events are characterized by a relatively rapid decrease in
the count rate followed by a more gradual return to the baseline level; that is,
the typical event profile is right-skewed. This pattern can be seen in
\autoref{typicalevents}, which shows four representative events from
\cite{chang2007}.

\begin{figure}
  \epsscale{1}
  \plotone{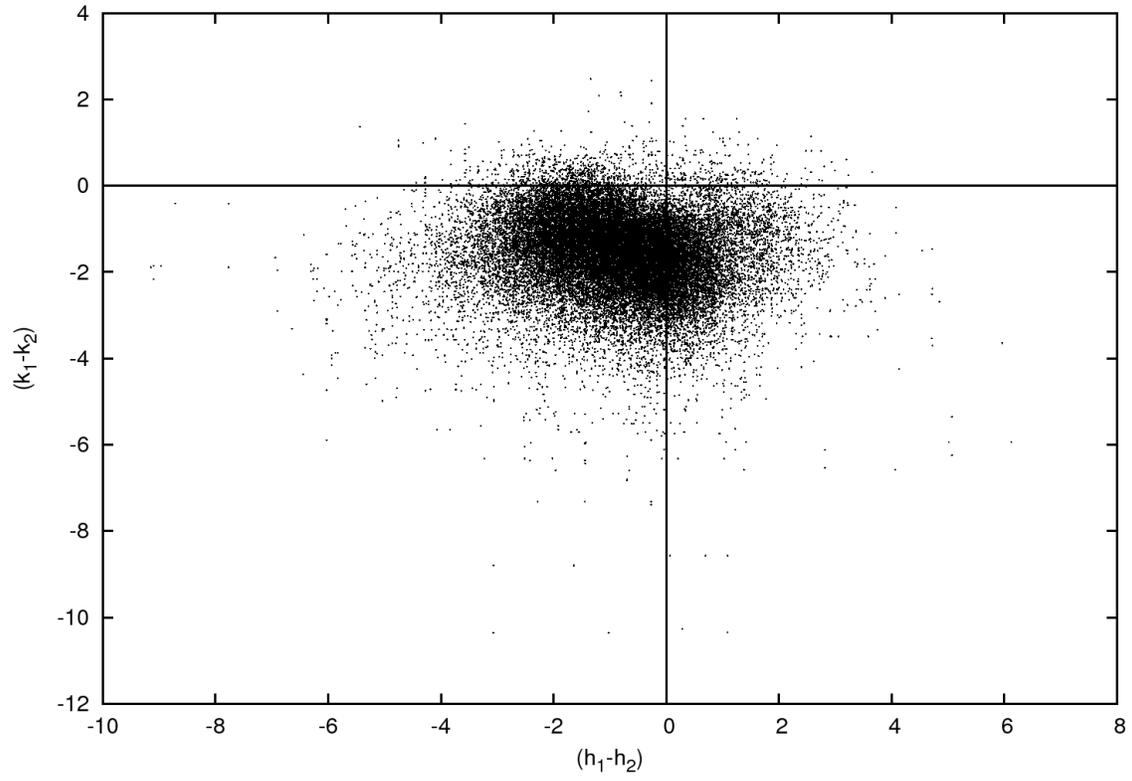}
  \caption{$(k_1-k_2)$ vs. $(h_1-h_2)$ for event 90
   from \cite{chang2007}.
   \label{event90posteriorscatter}}
\end{figure}

It is interesting to note that ($h_1-h_2$) and ($k_1-k_2$) are typically
negatively correlated (when examining their joint posterior distribution) for
approximately symmetric events. An example of this comes from our analysis of
event 90 from \cite{chang2007}. The joint posterior distribution of ($h_1-h_2$)
and ($k_1-k_2$) can be seen (as a scatterplot) in
\autoref{event90posteriorscatter}. Two points are immediately apparent from this
figure. First, the deviation of the distribution from the origin indicates that
this event is most likely asymmetric. Second, ($h_1-h_2$) and ($k_1-k_2$) are
strongly negatively correlated ($r = -0.618$). The intuition behind this outcome
is clear. We are allowing for two possible deviations from symmetry in our
model: deviations based on the half maxima ($h_1-h_2$) and deviations based on
the rate of change ($k_1-k_2$) in our source's intensity. For very short events,
we cannot be certain to which of these deviations we should attribute the
asymmetry of our event, as we can, to some extent, trade-off between them. Our
posterior distribution simply reflects this trade-off.

It is also important to note that, when analyzed using our method, the events
identified as potentially non-instrumental in origin by \citeauthor{chang2007}
do not appear to deviate significantly from the remaining events in FWHM, half
maxima deviation, or curvature deviation. This supports the conclusion of
\cite{liu2008mnras} that the events detected thus far (including those
identified as possibly non-instrumental) are most likely the result of dead-time
effects and other instrumental contamination, not TNOs occulting Scorpius X-1.

At the distance of the Kuiper Belt ($\sim 40$AU) the size of the objects given
the width of the events is  $\sim 50$ m, and that is close to the {\em Fresnel
scale}, which is $\sqrt{ \lambda \; a / 2}$, where $\lambda$ is the wavelength
and $d$ is the distance. We use   $\lambda = 0.3$ nm (4 keV) because most of the
RXTE/PCA-detected photons from Sco ~X-1 are at this energy. As a  result these
occultation events are diffraction dominated phenomena (see \cite{roques1987}
and \cite{niheietal2007}) and we expect that diffraction effects would produce
small count rate increases on either side of the dips.

\begin{figure}
 \epsscale{1}
 \plotone{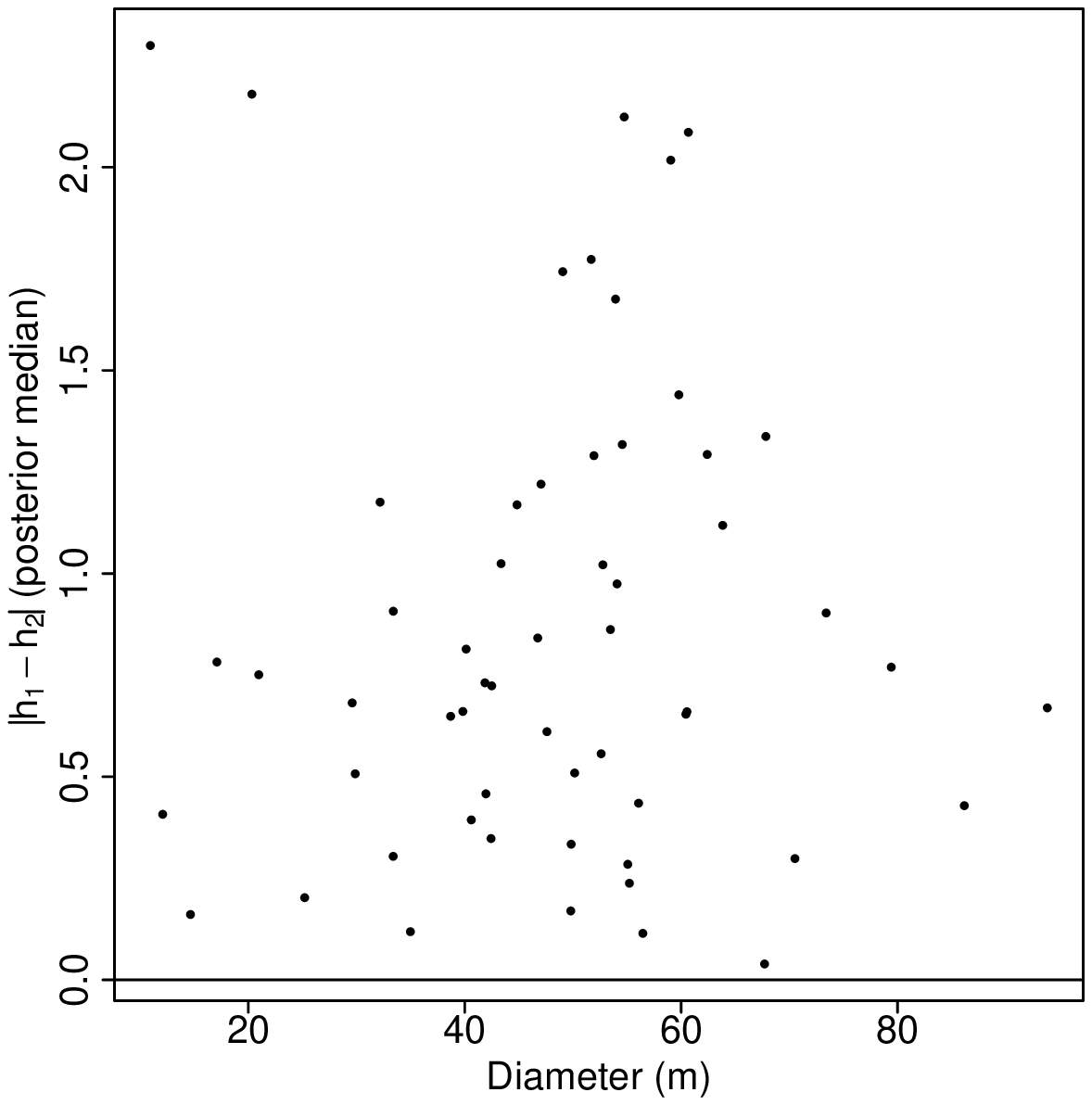}
 \caption{$|h_1-h_2|$ vs. estimated diameter for 57 of the 58 events from
 \cite{chang2006oxr} (58th event omitted due to anomalous nature).
 \label{abshdiffvssize}}
\end{figure}

\begin{figure}
 \epsscale{1}
 \plotone{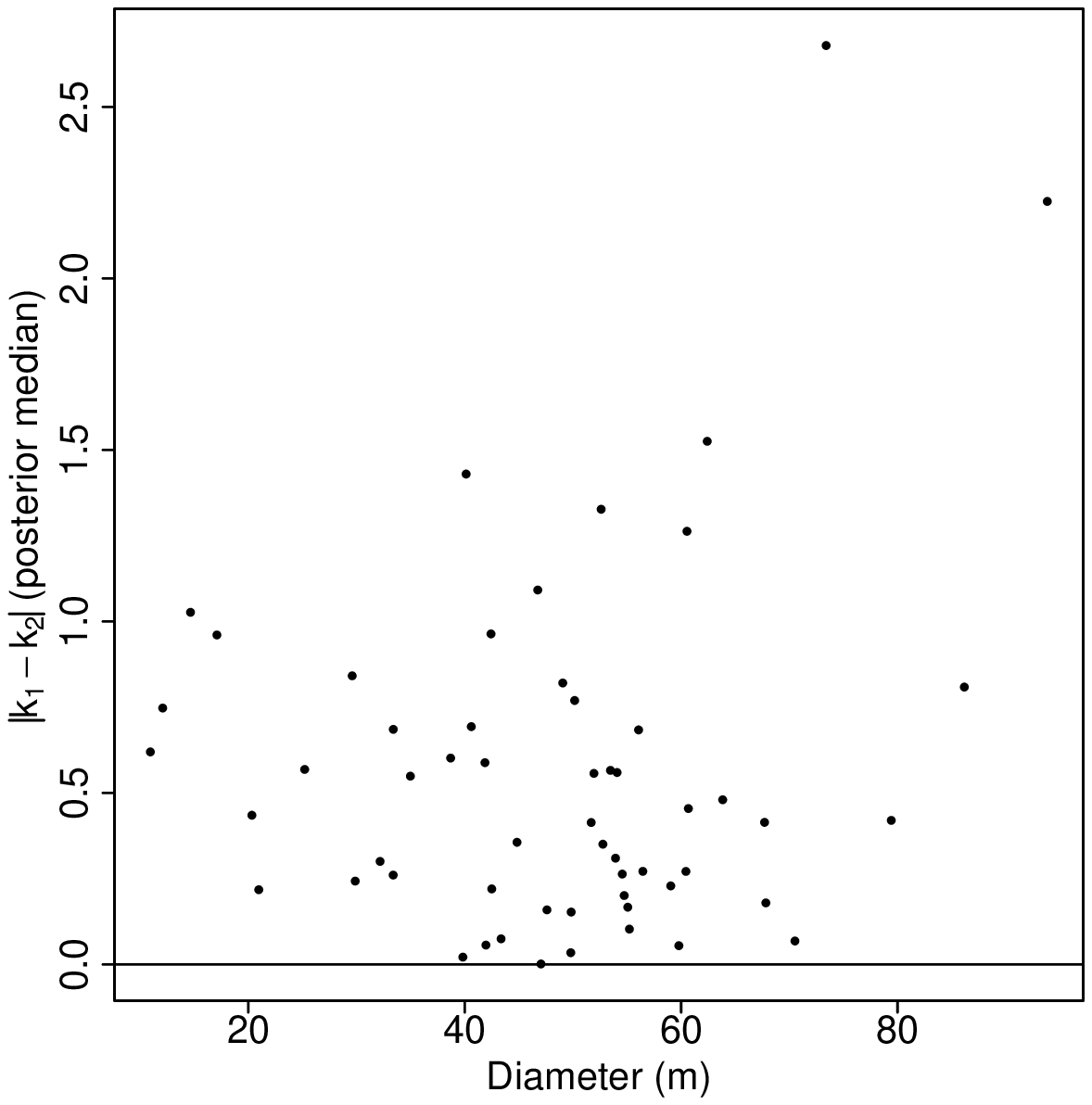}
 \caption{$|k_1-k_2|$ vs. estimated diameter for 57 of the 58 events from
 \cite{chang2006oxr} (58th event omitted due to anomalous nature).
 \label{abskdiffvssize}}
\end{figure}

Also TNOs of this size are not expected to be spherically symmetric
\citep{roques1987} and therefore the resulting occultation events  in the
light curves  are  not expected to be time symmetric. However for TNOs of the
size of the Fresnel scale or smaller  the shape of the object does not translate
into  asymmetry of the time series. Roques et al 1986 has studied light curves
from different shaped TNOs and demonstrated that when the effective size of the
TNOs is smaller than the Fresnel scale the resulting events in the light
curve do not differ from the light curves signatures  of regularly spherically
shaped TNOs. The sizes of the 107 objects studied here are within the range of
the Fresnel scale so some asymmetry on the light curves should be manifested.
Figures \hyperref[abshdiffvssize]{\ref*{abshdiffvssize}} and
\hyperref[abskdiffvssize]{\ref*{abskdiffvssize}} show the measures of asymmetry
as a function of the size of the object for 57 of the 58 events from
\cite{chang2006oxr}. As can be seen from the figures there is no clear relation
between the estimated size of an object and the level of asymmetry for its
corresponding event. This provides further evidence that the observed dips in
intensity do not correspond to TNO transit events, as \cite{jones2008apj}
suggest.

\section{Conclusions}
\label{sec:conclusions}

Based on our analysis of the suspected Scorpius X-1 occultation events, we
conclude that there is significant evidence for temporal asymmetry in these
events. This finding is consistent with the suggestion of
\citet{jones2008apj} that the observed dips in brightness are due to dead time
effects. We also conclude that the events identified in \cite{liu2008mnras} as
possibly noninstrumental in origin are most likely not occultations.

We also note that the approach presented herein can be applied (with minor
modifications) to the study of events in nearly any light curve. Such studies
need not be limited to examining the symmetry of events, as the
parameterization of our bilogistic event profile and flexible Bayesian approach
allow inference to be made on quantities such as full width at half maxima. For
example, our approach could be extended to analysis of exoplanetary transits,
supernovae, quasars or any other transient event.

Furthermore, the method we have presented for constructing a proposal
distribution for the center of each event can be applied quite generally. Even
when using a physically grounded model with parameters such as body size, angle
of incidence, etc., one must account for uncertainties about the location of the
event in the light curve when making inferences about the parameters of
interest. This can be quite challenging using classical methods, and even MCMC
methods do not offer an panacea. Our approach to constructing an independence
chain proposal distribution for the location parameter in such models makes them
quite a bit more tractable, especially in the case of multiple events.

\section*{Acknowledgements}
A.W.B., P.P., and C.A. gratefully acknowledge support from NSF IIS-0713273. We
would also like to thank the participants in the Harvard Astrostatistics Seminar
for their feedback, particularly Xiao-Li Meng. The simulations in this paper
were run on the Odyssey cluster supported by the Harvard FAS Research Computing
Group.

\bibliography{sco1timesymmetry} 

%

\end{document}